\documentclass [12pt]{article}
\title{COSMOLOGICAL CONSTANT, QUINTESSENCE AND EXPANSIVE
  NONDECELERATIVE UNIVERSE}

\author{Jozef \v{S}ima and Miroslav S\'{u}ken\'{\i}k\\[1ex]
  Slovak Technical University, Radlinsk\'{e}ho 9, \\
  812 37 Bratislava, Slovakia}

\date{}

\begin{document}
\maketitle

\begin{abstract}
  Recent observations of the Universe have led to a conclusion
  suppressing an up-to-now supposed deceleration of the Universe
  caused by attractive gravitational forces. Contrary, there is a
  renaissance of the cosmological member $\Lambda $ and introduction
  of enigmatic repulsive dark energy in attempts to rationalize a
  would-be acceleration of the Universe expansion. It is documented
  that the model of Expansive Nondecelerative Universe is capable to
  offer acceptable answers to the questions on the Universe expansion,
  state equations of the Universe, the parameter $\omega $, the
  cosmological member $\Lambda $ without any necessity to introduce
  new strange kinds of matter or energy being in accord with the
  fundamental conservation laws and generally accepted parameters of
  the Universe.
\end{abstract}

\bigskip

\section{Background of the ENU model}

Solving Einstein equations of field by means of Robertson-Walker
metric, Friedmann obtained equations of universe dynamics. In the
Expansive Nondecelerative Universe (ENU) model it has been
rationalized [1, 2] that the Universe, throughout the whole expansive
evolutionary phase, expands by the velocity of light $c$, and the
gauge factor $a$ can be thus expressed as
\begin{equation}
\label{eq1}
a = c.t_{c} 
\end{equation}
where $t_{c} $ is the cosmological time (the present values provided by 
calculations based on ENU are $a \cong 1.3x10^{26}$m; $t_{c} \cong 
4.3x10^{17}$s). In this approach, the curvature index $k$ and cosmologic 
member $\Lambda $ are of zero value
\begin{eqnarray}
k & = &0 \label{eq:2a} \\
\Lambda &=&0
\end{eqnarray}
Solving Friedmann equations in the ENU leads to 
\begin{equation}
\label{eq2}
c^{2} = {\frac{{8\pi .G.\rho .a^{2}}}{{3}}} = - {\frac{{8\pi 
.G.p.a^{2}}}{{c^{2}}}}
\end{equation}
where $\rho $ is the mean (critical) mass density of the
Universe (at present, $\rho \approx 10^{ - 26}$ kg m$^{{\rm -} {\rm
    3}}$) and $p$ is the pressure.  Since the energy density
$\varepsilon $ is
\begin{equation}
\varepsilon = \rho .c^2
\end{equation}
in accordance with the General Theory of Relativity and (\ref{eq2}),
the Universe with total zero and local non-zero energy meets the state
equation [3]
\begin{equation}
\label{eq3}
p = - {\frac{{\varepsilon} }{{3}}}
\end{equation}

Providing that the velocity of the Universe expansion is equal to $c$, the 
total gravitational force must be equal to zero (compare 1 and 6). Equation 
(\ref{eq2}) leads then to the expression for $\rho $

\begin{equation}
\label{eq4}
\rho = {\frac{{3c^{2}}}{{8\pi Ga^{2}}}}
\end{equation}
and, due to~(\ref{eq:2a}), for $\rho $ representing the density of the
Universe with the mass $m_{u}$, it must be valid for a flat Universe
at the same time

\begin{equation}
\label{eq5}
\rho = {\frac{{3m_{u}} }{{4\pi .a^{3}}}}
\end{equation}

Taking into account the expressions~(\ref{eq4}) and~(\ref{eq5}), for
the gravitational radius it then holds
\begin{equation}
\label{eq6}
a = {\frac{{2G.m_{u}} }{{c^{2}}}}
\end{equation}

Since $a$ is increasing in time, $m_{u}$ must increase as well (its
present value approaches 8.6 $\times $ 10$^{{\rm 5}{\rm 2}}$ kg),
$i.e.$ in the ENU, the creation of matter occurs~[3]. The total energy
of the Universe must, however, be exactly zero. It is achieved by a
simultaneous gravitational field creation, the energy of which is
$E<0$. The fundamental mass-energy conservation law is thus observed.
Due to the matter creation, in ENU Schwarzschild metrics must be
replaced by Vaidya metrics [4, 5].

The ENU model has been up to now able to offer answers to several
current questions of both micro-world and micro-world [6]. In
connection to the matter discussed in this contribution, we would like
to remind the zero value of $\Lambda $ and the state
equation~(\ref{eq3}) leading directly to $\omega $ value
\begin{equation}
\label{eq7}
\omega = - 1 / 3
\end{equation}
where the parameter $\omega $ is defined as the ratio
\begin{equation}
\omega = p/\varepsilon 
\end{equation}

From the viewpoint of ENU, gravitation is a local phenomenon which can
manifests itself only in cases when the total energy density exceeds
the critical energy density. The total average energy density of the
Universe is identical to the critical density and thus, from the
global point of view, gravitational forces cannot exhibit themselves.

A typical nature of the ENU model is its simplicity, no ``additional
parameters'' or strange ``dark energy'' needed, and the usage of only
one state equation in describing the Universe. Calculated gauge factor
$a$, cosmological time $t_{c} $, and energy density $\rho $ match well
the generally accepted values.

\section{Cosmological member $\Lambda$}

One of the corner stones of many modern cosmological theories has for
long time been a postulate stating that after its inflation period,
the Universe has undergone a classic Friedman expansion which would
gradually decelerate due to the gravitational force influence. The
measure of this deceleration would depend on the energy density of the
Universe.

Relation between the pressure $p$ and the energy density
$\varepsilon $ of the Universe can be expressed via three
equations, namely:
\begin{eqnarray}
p & = &\varepsilon \label{eq:12a}\\
p & = &\varepsilon/3 \label{eq:13a}\\
p & = &0 \label{eq:14a}
\end{eqnarray}
Relation~(\ref{eq:12a}) describes the relativistic particles,
equation~(\ref{eq:13a}) which is used for the radiation
era,~(\ref{eq:14a}) is the state equation of dust particles and
can be applied in the matter era.

Measurements of radiation intensity (brightness) of distant supernovas
performed in the last years have shown that this intensity is
significantly lower than that predicted. Consequently, these objects
should be more distant than expected and, in turn, there should be no
deceleration in the Universe expansion caused by gravitation. It seems
that these observations justify the introduction of the cosmological
member $\Lambda $. This member represents the negative
pressure or repulsive forces of the vacuum counteracting to the
attractive forces of gravitation. The cosmological member was
introduced by Einstein as a mode of the preservation of static
character of the Universe.

Stemming from the Hubble discovery and characterization of the
Universe expansion, there is no need to take the cosmological member
into account any longer.

The second introduction of $\Lambda $ was connected to an
inaccuracy in the Hubble constant $H$ determination. Too high value
having been estimated would led, however, directly to a too short age
of the Universe. The problem was overcome introducing $\Lambda
  $ which allowed to prolong the age. After performing a more precise
determination of $H$, the cosmological member become irrelevant.

The third instance of the cosmological member $\Lambda $
introduction into theory appeared in relation to the inflation model
of the Universe, where this member describes certain quantum effects of
the physical vacuum.

\section{Parameter $\omega$}

For cases of a nonzero value of the cosmological member $\Lambda $ it must 
hold
\begin{equation} \label{eq:15a}
\omega = -1
\end{equation}
Owing to some generally accepted present parameters of the Universe
(such as gauge factor, cosmological time, critical energy density)
many problems appear in the models of the Universe incorporating the
cosmological member.  In the inflation model with a nonzero
cosmological member, the above parameters values can be approached in
such a way that following the inflation period the Universe
decelerated due to gravitation and matter emerged beyond the causal
horison first, and then due to an influence of the cosmological
member, the Universe expansion accelerated. To describe such a
Universe, two state equations are required as a minimum. One can,
however, hardly believe that just at the time being the critical
energy density of the Universe is reached. Moreover, it is not clear
how this critical energy density would be kept during a gradual
acceleration of the Universe expansion in the future. The most recent
measurements of $\omega $ suggest that the value
\begin{equation}\label{eq:16a}
\omega = -2/3
\end{equation}
is more probable than the value in~(\ref{eq:15a}).

\section{The fifth force}

A newly established model of the Universe -- quintessential model [7]
-- is based on the supposed existence of the fifth force, i.e. a
special dark energy, which generates negative pressure and thus
contributes to the acceleration of the Universe expansion. This
quintessential energy is not so ``rigid'' as the cosmological member
is, contrary, it is ``soft'' and its properties and effects can vary
during the Universe expansion.

Quintessential model works with a value of $\omega $ in the range
\begin{equation}
\label{eq8}
\omega = ( - 1, - 1 / 3)
\end{equation}
the most probable being the value given by equation~(\ref{eq:16a}).
The inherent problems of this model lie in the unability to
characterize the matter responsible for the creation of the
quintessential energy. In this model several state equations are used
and it is highly improbable that a combination of such state equations
could lead just to the present critical energy density.

\section{Conclusions}

The corner stones of the model of Expansive Nondecelerative Universe
(ENU) were laid in the nineties of the last century [2]. It was shown
[3] that in the ENU model $\omega = - 1 / 3$ (\ref{eq7}), which is a
value falling into a range predicted based on the latest experimental
observations and calculations.

To determine an exact value of $\omega $, several new experimental
observations have been proposed, e.g. utilization of a satellite SAP
(Supernova Acceleration Probe) to investigate distant supernovas, and
an Earth-located telescope ELASST able to measure the differences in
angle dimensions of hot and cold spots in microwave radiation of the
cosmic background. It should be pointed out, however, that in
evaluation of $\omega $ value, its interrelation with Hubble constant
must be taken into account. In the case of SAP project, the value of
Hubble constant must be determined as precisely as possible. A lower
value of $H$ could prefer the ENU model with $\omega $ value given in
(\ref{eq7}), a higher value would lead to quintessential model
with~(\ref{eq:16a}) or, vice versa, the measured results can be
rationalized using (\ref{eq7}) for a lower $H$ or~(\ref{eq:16a}) for a
higher value of $H$.

\section{References}

V. Skalsk\'{y}, M. S\'{u}ken\'{\i}k, Astrophys. Space Sci., 178 (1991) 169

J. \v{S}ima, M. S\'{u}ken\'{\i}k, Preprint gr-qc/9903090

V. Skalsk\'{y}, M. S\'{u}ken\'{\i}k, Astrophys. Space Sci., 209 (1993) 123

P. Vaidya, Proc. Indian Acad. Sci., A33 (1951) 264

M. S\'{u}ken\'{\i}k, J. \v{S}ima, Astrophys. Space Sci., submitted

M. S\'{u}ken\'{\i}k, J. \v{S}ima, Preprint gr-qc/0010061

G. Huey, J.E. Lidsey, Preprint astro-ph/0104006

\end{document}